\documentstyle[epsfig,osa,twocolumn]{revtex}

\begin{document}

\title{The Decay of Passive Scalars Under the Action of Single Scale Smooth Velocity Fields in 
Bounded 2D Domains : From {\it non} self
similar pdf's to self similar eigenmodes.}
\author{Jai Sukhatme and Raymond T. Pierrehumbert}
\address{Dept. of Geophysical Sciences, University of Chicago, Chicago
, IL 60637}
\date{\today}
\maketitle

\begin{abstract} 
We examine the decay of passive scalars with small, but non zero, diffusivity in 
bounded 2D domains. 
The velocity fields responsible for advection are smooth (i.e., they
have bounded gradients) and of a single large scale. Moreover,
the scale of the velocity field is taken to be similar to the size of the entire domain.
The importance of the initial scale of variation of the scalar field with respect to that of the
velocity field is strongly emphasized. 
If these scales are comparable and
the velocity field is time periodic, we see the formation of a periodic 
scalar eigenmode. The eigenmode is numerically 
realized by means of a deterministic 2D map on a lattice. Analytical justification
for the eigenmode is available from theorems in the dynamo literature.
Weakening the notion of an eigenmode to mean statistical stationarity,
we provide numerical evidence that the eigenmode solution also holds for aperiodic flows
(represented by random maps).
Turning to the evolution of an initially small scale scalar field,
we demonstrate the transition from an evolving (i.e., {\it non}
self similar) pdf to a stationary (self similar)
pdf as the scale of variation of the scalar field progresses from being 
small to being comparable to that of the velocity field (and of the domain). 
Furthermore, the {\it non} self similar regime itself consists of two stages. 
Both the stages are examined and the coupling between diffusion and the distribution
of the Finite Time Lyapunov Exponents is shown to be responsible for the pdf
evolution. 
\end{abstract}


\narrowtext

\section{Introduction} 

Starting with the work of Batchelor \cite{Batch-59} the study of passive
scalars in smooth velocity fields has been the subject of numerous investigations.
Originally posed in 3D \cite{Batch-59}, the problem considered a situation where the viscosity
($\nu$) of the flow is much greater than the diffusivity ($\kappa$) of the 
scalar. Therefore, even at large Reynolds numbers, for length scales 
below the viscous cutoff ($l_{\nu}$) and
above the diffusive cutoff ($l_{\kappa}$), the so called Batchelor regime,
one has a smooth velocity advecting a diffusive tracer. Since then the problem 
has grown to encompass a variety of phenomena such as chaotic advection (see for eg.
Ottino \cite{Ottino}) and scalars in inverse cascading 2D turbulent flows
(see Section III in Falkovich et. al. \cite{Falk} for a recent review). 
Of particular interest to us are applications in the realm of geophysical fluid dynamics,
specifically,
the mixing of scalars along isentropic surfaces via large scale atmospheric flows \cite{Ray-JAS},\cite{Hu}. 
\\

The equation governing the passive scalar ($\phi$) is,

\begin{equation}
\frac{\partial \phi}{\partial t} + (\vec{u} \cdot \nabla) \phi = \kappa \nabla^2 \phi
\label{1a}
\end{equation}
Let us denote the size of the domain by $L$.
This is a linear equation for $\phi$, 
the velocity field is part of the 
prescribed data. The usual conditions imposed on the velocity field are that it be
divergence free and smooth, i.e.,
$\nabla \cdot \vec{u} = 0$ and $|\nabla \vec{u}| < \infty$ everywhere in the domain. 
The questions asked are usually of the form : 
given an initial scalar field is it possible to determine the 
long time behaviour of the solution ? Does one see the emergence of an eigenmode with a well 
defined decay rate, in other words what is the nature of the spectrum of the advection
diffusion operator in a suitably defined function space ? 
Usually, given the aperiodic nature of the 
advecting velocity fields the question is better posed in a statistical sense, i.e., is the 
asymptotic pdf of the scalar field self similar or is it an ever evolving ({\it non} self similar)
entity ? 
As things stand,
the results found in the literature are fairly divided. 
One of the aims of this paper is to unify these results
by bringing out certain salient features (such as scale separation) 
which have not been emphasized previously.
\\

Most theoretical studies have focused on the case when 
the initial scale of variation of the scalar field (say $l_s$, $l_s >> l_{\kappa}$)
is much smaller than that of the velocity field (say $l_v$, we take
$l_v \sim L$). The
problem then is to determine the statistical properties of the scalar 
field at scales between $l_v$ and $l_{\kappa}$. 
A successful approach in this direction has been to shift to a comoving
reference frame, use the decomposition
$u_\alpha = (\partial_{\alpha} u_\beta) (t) ~ r_\beta = \sigma_{\alpha \beta}(t) ~ r_\beta$ 
and deal 
with the
the effective equation that results from a substitution in Eq. (\ref{1a}) 
(see for eg. \cite{Chert-96}, \cite{BF-99} 
and the references therein). 
The expectation \cite{BF-99} is that the scalar field will have {\it non} self similar pdf's. 
Or, defining the moments of the scalar field as $< |\phi(x,y,t)|^n >$, Balkovsky and
Fouxon \cite{BF-99} explicitly show that,

\begin{equation}
< |\phi(x,y,t)|^n > \sim e^{-{\alpha}_n t} ~;~ t> T
\label{1b}
\end{equation}
and that ${\alpha}_n$ is a nonlinear function of $n$ ($T$ is a diffusive time
scale which will be clarified later), implying the 
{\it non} self similarity of the pdf's. 
Recent experiments \cite{Jul} with passive scalars 
injected at point sources in 
inverse cascading $2$D turbulent flows seem to validate some of these predictions. 
Concrete evidence is available from numerical work demonstrating the nonlinearity of
$\alpha_n$ for scalars advected by realistic atmospheric winds \cite{Hu}. It should be noted that 
these simulations are over a time scale which represent the transient scalar behaviour
rather than the asymptotic large time solution.
\\

On the other hand, there is a small body of work which deals with the case when
$l_{\kappa} << l_s \sim l_v \sim L$, i.e., the initial scale of variation of the 
scalar field is comparable to that of the velocity field, both of which are in
turn comparable to the scale of the domain.
\footnote{We always take $l_v \sim L$. There are other problems of
interest where $l_v << L$ which are studied in the
context of turbulent diffusion (see for eg. the comprehensive review by
Majda and Kramer \cite{Maj}).\\}
The 
observation, based on numerical work \cite{Ray-94},\cite{Ray-Chaos}, 
is that the scalar field enters an
eigenmode (termed a "strange eigenmode" by Pierrehumbert \cite{Ray-94} due to its 
spatial complexity). The pdf of the scalar becomes self similar (after
a suitable normalization by the variance) or in terms of the moments, $\alpha_n \sim n$. 
Recently experimental evidence for 
the emergence of such an eigenmode has been provided in the case of time 
periodic velocity fields \cite{Gol-Nat}. A line of attack on this problem has been to 
try and relate the Finite Time Lyapunov Exponents (FTLE's) of the advecting flow to the
statistical properties of the scalar field (see \cite{Ant} and the references therein). 
Even though this work \cite{Ant} is successful in describing the initial stages of the 
problem, the FTLE's are by definition local entities
and appear to be unsuitable precisely when $l_s \sim l_v$
\footnote{
{\it Note} : A recent paper \cite{Won} analyzes
the work of Antonsen et. al \cite{Ant} with an aim to understand the exponential decay
of scalar moments. This paper, Wonhas and Vassilicos \cite{Won},
identifies a "global" mechanism
responsible for the exponential decay and also hints at the importance of the
scale separation we have emphasized in this paper.
}.
\\

In this study we work with simple prescribed velocity fields in 
$2$D that have a single large scale (of the 
kind encountered in studies of chaotic advection). It is worth emphasizing that the 
power spectrum of such velocity fields is discrete. This is in contrast to 
technically smooth velocity fields which possess a heirarchy of scales.
In other words the velocity fields considered are Lipschitz continuous over
all scales of interest (i.e., between $l_\kappa$ and $l_v \sim L$).
The large single scale, divergence free and smooth nature of the 
velocity field implies that the trajectory equations do not show explosive (implosive) 
separation (collapse) as is generally expected in turbulent (or multiple scale) incompressible 
(compressible) velocity fields
(see for eg. Gawedzki and Vergassola \cite{Gaw}).
\\

Our first aim is to give further justification for the eigenmode 
in periodic and aperiodic velocity fields. We explicitly construct an area
preserving deterministic  $2$D map
which, in spite of having weak barriers, demonstrates the emergence of a periodic
eigenmode. 
Moreover, we point out theorems in the dynamo literature which are applicable
to the passive scalar problem, thereby providing analytical justification 
(in terms of the purely discrete spectrum of the advection diffusion operator) for the eigenmodes.
The aperiodic case is free of barriers due to the random nature of the maps
employed but is slightly novel
in the sense that the notion of an
eigenmode has to be weakened by invoking statistical stationarity (i.e. self similarity of the pdf).
Secondly, we present a 
unified picture of the two seemingly disparate cases mentioned above. 
Starting with a 
scalar field which satisfies $l_s << l_v \sim L$, we observe an evolution in the shape of the pdf (i.e., 
the pdf is {\it non} self similar). 
This 
evolution ceases when the scales of the velocity and scalar fields become comparable
(i.e., the field enters an eigenmode with a self similar pdf). The various regimes that the
scalar field encounters are analyzed and a simple
example, motivated by the work of Balkovsky and Fouxon \cite{BF-99}, is used to 
elucidate the coupling between diffusion and the distribution of FTLE's and its role in 
determining the pdf evolution. 
A discussion of the regimes, their universality and the issue of 
scale separation concludes
the paper.
\\

\section{The Eigenmode}
\subsection{The Eigenvalue Problem}

Physically when $l_s \sim l_v \sim L$ the scalar field feels the effect of the finite
size of the domain,
moreover, as $l_s \sim l_v$ it is difficult to justify any linearization of the
underlying nonlinear trajectory equations. Fortunately, the linear 
nature of Eq. (\ref{1a}) lends itself to an eigenvalue analysis. Recently 
Fereday et. al \cite{Fer} have looked at the eigenvalue problem for a specific
velocity field (represented by the Bakers map).
We consider the eigenvalue problem in its full generality as has been 
done for the magnetic field in the kinematic dynamo literature \cite{Bayly1}, \cite{Bayly2}, \cite{Sow-94}. 
Most of the material in this subsection is well known in the dynamo literature, we 
present it here merely to put things in a well defined framework. 
Let us recast Eq. (\ref{1a}) as,

\begin{equation}
\frac{\partial \phi}{\partial t} = {\mathcal L}_{\kappa} \phi \quad ; \quad
{\mathcal L}_{\kappa} \phi = \kappa \nabla^2 \phi - (\vec{u} \cdot \nabla) \phi
\label{1c}
\end{equation}
For a steady velocity field one can separate time by assuming a solution of the 
form $\phi(x,y,t) = \overline{\phi}(x,y) e^{\sigma t}$. Hence the eigenvalue problem is,

\begin{equation}
{\mathcal L}_{\kappa} \overline{\phi} = \sigma \overline{\phi}
\label{1d}
\end{equation}
Here $\overline{\phi} \in {\it B}({\sf D})$, where ${\it B}({\sf D})$ is a Banach space
of square integrable functions 
\footnote {For {\it finite} times, given that $|\nabla \vec{u}| < \infty$, we are
guaranteed that the scalar field will remain square integrable.\\}
defined on the domain {\sf D}. We take {\sf D} to be
$[0,2\pi] \times [0,2\pi]$ with opposite sides identified (i.e., a torus). 
The object of interest is the spectrum of the operator ${\mathcal L}_{\kappa}$ acting
in {\it B}({\sf D}). As we are working in a Banach space of infinite dimensions
the spectrum 
depends strongly on the nature of the operator \cite{Kato}. 
\\

If the operator in question were compact \footnote{Consider an infinite sequence of 
functions $\psi_1, \psi_2...$ in a complete function space. Let us denote the 
action of an operator $A$ on these functions by the sequence 
$A \psi_1, A \psi_2...$. If $A$ is closed with respect to this space and further 
if one can extract a convergent subsequence from
$A \psi_1, A \psi_2...$ then $A$ is 
called a compact operator.\\} 
then the spectral theory of finite dimensional operators
would carry over to our infinite dimensional problem \cite{Kato}.
Unfortunately differential operators are usually not compact, moreover they
are unbounded. One of the techniques used to handle such operators is 
to gain control over their resolvents (see Kato \cite{Kato} Chap. 3), the
case when the operator is closed and its resolvent is compact is particularly
clear cut \cite{Kato}.
As it turns out this is precisely what happens to ${\mathcal L}_{\kappa}$
for steady velocity fields and non zero $\kappa$, i.e., it
is (weakly) closed on ${\it B}({\sf D})$
and has a compact resolvent \cite{Child-book}.
\footnote{The cited reference \cite{Child-book} deals with a similar operator
which appears in the magnetic dynamo theory, specifically the operator they
work with is,

\begin{equation}
{\mathcal M}_{\kappa} \vec{B} = \kappa \nabla^2 \vec{B} - (\vec{u} \cdot \nabla) \vec{B}
+ \vec{B} \cdot \nabla \vec{u}
\label{1e}
\end{equation}
where $\vec{B}$ is the magnetic field. The difference in ${\mathcal L}_{\kappa}$ and
${\mathcal M}_{\kappa}$ is the additional "stretching" term in ${\mathcal M}_{\kappa}$.
As this term does not contain derivatives of $\vec{B}$ it is quite intuitive that
the spectral properties of ${\mathcal M}_{\kappa}$ should carry over to ${\mathcal L}_{\kappa}$.
In fact, it is quite reasonable that as long as the gradients of the velocity field are bounded,
the spectral properties of both ${\mathcal L}_{\kappa}$ and
${\mathcal M}_{\kappa}$ follow those of the Laplacian as the highest order
derivatives are contained in the Laplacian term.\\
}
Hence its spectrum consists of a discrete set of eigenvalues of 
finite multiplicity (Kato \cite{Kato} p. 187), in fact it has a complete
set of eigenfunctions \cite{Child-book}. Physically the implication of a 
purely discrete spectrum is that the eigenvalues are well separated, as time
goes on $\overline{\phi}(x,y)$ will assume the form of the eigenfunction 
corresponding to the largest eigenvalue.  
In the more general case of a periodic
velocity field we have to consider the Floquet problem involving the propagation of 
$\phi(x,y,t)$ to $\phi(x,y,t+T)$ where $T$ is the periodicity of the underlying flow.
Denoting the propagator by ${\mathcal T}_{\kappa}(T)$, 
the eigenvalue problem is of the form,

\begin{equation}
{\mathcal T}_{\kappa}(T) \overline{\phi} = e^{pT} ~ \overline{\phi}
\label{1f}
\end{equation}
The discreteness of the spectrum of ${\mathcal T}_{\kappa}$ persists 
but the existence of a complete set of eigenfunctions is not
gauranteed \cite{Child-book}. 
Essentially when the flow is smooth and $\kappa > 0$, 
we are assuming that the spectral properties of ${\mathcal L}_{\kappa}$
follow those of ${\mathcal M}_{\kappa}$ (Eq. (\ref{1e})) for which there
exist rigorous results in the cited dynamo literature.
So, from a mathematical
point of view the problem has a purely
discrete set of eigenvalues and a (possibly incomplete) set of eigenfunctions
for both steady and time periodic velocity fields. In passing we mention
that the $\kappa \rightarrow 0$ limit is expected to be quite delicate;
interesting aspects not encountered in the 
$\kappa \neq 0$ problem are likely to appear in this limiting process
(see Bayly \cite{Bayly2} for some illuminating examples in regard to the analogous scalar
dynamo problem). 
\\

\subsection{Deterministic Maps (Periodic Velocity Fields)}
\begin{figure}
\begin{center}
\epsfxsize=9.0 cm
\epsfysize=9.0 cm
\leavevmode\epsfbox{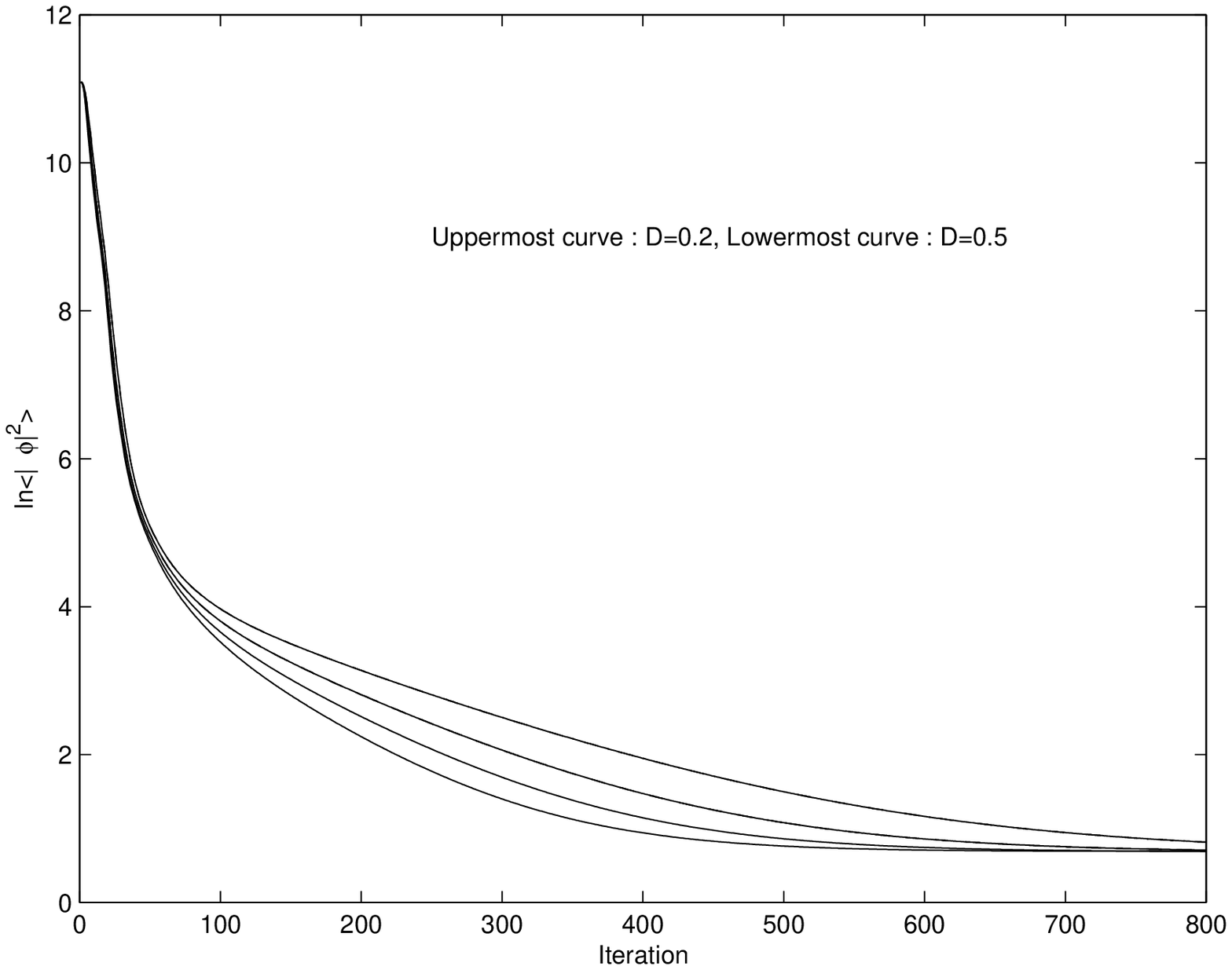}
\end{center}
\caption{The strong barrier case : The decay of the variance for different diffusivities.}
\label{fig:fig1}
\end{figure}

We proceed to see if we can numerically realize the eigenfunctions corresponding to the above
mentioned eigenvalues. 
Chaotic maps on a lattice are utilized to represent the 
advection process (see Pierrehumbert \cite{Ray-Chaos} for a succinct overview of lattice maps in advection). 
The advantage of the lattice maps is the exact preservation
of scalar moments and their numerical efficiency. This "pulsed" advection is 
followed by a diffusive step (for eg. see \cite{Ray-Chaos} or \cite{Ray-94}). 
Usually maps with random phase shifts
at each iteration are employed so as to break the KAM tori which are generically present in
deterministic area preserving maps. 
As this procedure gives an aperiodic time dependence to the flow it is 
unacceptable in the eigenvalue formulation outlined in the 
previous section.
Unfortunately, we are unaware of any area preserving, nonlinear and continuous map 
which has been proven to be mixing \footnote{Mixing in the sense of dynamical
systems (see for eg. \cite{Aref-91}).\\} over the whole domain {\sf D}.
The only
way we have of seeing that there are no regions where the scalar field remains
trapped is to do so numerically, i.e., to examine the variance as a function of time. 
The particular map we use mimics an alternating nonlinear shear flow,

\begin{eqnarray}
x_{n+1} = x_{n} + A_1 ~ \textrm{sin}(B_1 y_{n} + s_1) + A_2 ~ \textrm{sin}(B_2 y_{n} + s_1)  \nonumber \\
y_{n+1} = y_{n} + C_1 ~ \textrm{sin}(D_1 x_{n+1} + s_2) + \nonumber \\ 
~  C_2 ~ \textrm{sin}(D_2 x_{n+1} + s_2) 
\label{1g}
\end{eqnarray}
Here $s_1,s_2$ are constant random numbers used to get the sine functions out of 
phase and $x_{n},y_{n}$ are $\textrm{mod}(0,2\pi)$. The constants are 
$A_1 = 4, B_1 =1 , A_2 =1, B_2 = \pi , C_1=1, D_1=1, C_2=\frac{1}{4}, D_2=\frac{\pi}{2}$ and 
the initial condition of the 
scalar field is $\phi(x,y) = \textrm{cos}(x) \textrm{cos}(y)$ (note that 
the scale of the scalar
field is comparable to the scale of the flow). The diffusive step is implemented as,

\begin{equation}
\phi_{i,j} = (1-D)\phi_{i,j} + \Delta \phi_{i,j}
\label{1h}
\end{equation}
where $\Delta \phi_{i,j}$ is,

\begin{equation}
\Delta \phi_{i,j} = \frac{1}{4}D(\phi_{i+1,j}+\phi_{i-1,j}+\phi_{i,j+1}+\phi_{i,j-1})
\label{1i}
\end{equation}
Here $0 < D < 1$ is the diffusion coeffecient and $i,j$ are the indices of the 
lattice.
As is expected, for certain
combinations of $s_1,s_2$ there are persistent KAM barriers and we do not observe
a decay of the scalar variance as would be expected from a globally
mixing system. In fact, after a short time the diffusive exchange 
across the barriers controls the variance. 
$s_1 = 0 , s_2 = 0$ provides an example of this phenomenon, 
Fig. \ref{fig:fig1} shows the variance as a function of time for $D=0.5, 0.4, 0.3$ and
$D=0.2$, the control of the variance by the diffusion coefficient
is evident, moreover, a certain amount of scalar is trapped within the 
barriers (the trapping would be permanent as $\kappa \rightarrow 0$, 
though the scaling
of the amount that diffuses as a function of $\kappa$ is expected to be nontrivial \cite{Bern}).
\\

\begin{figure}
\begin{center}
\epsfxsize=12.0 cm
\epsfysize=15.0 cm
\leavevmode\epsfbox{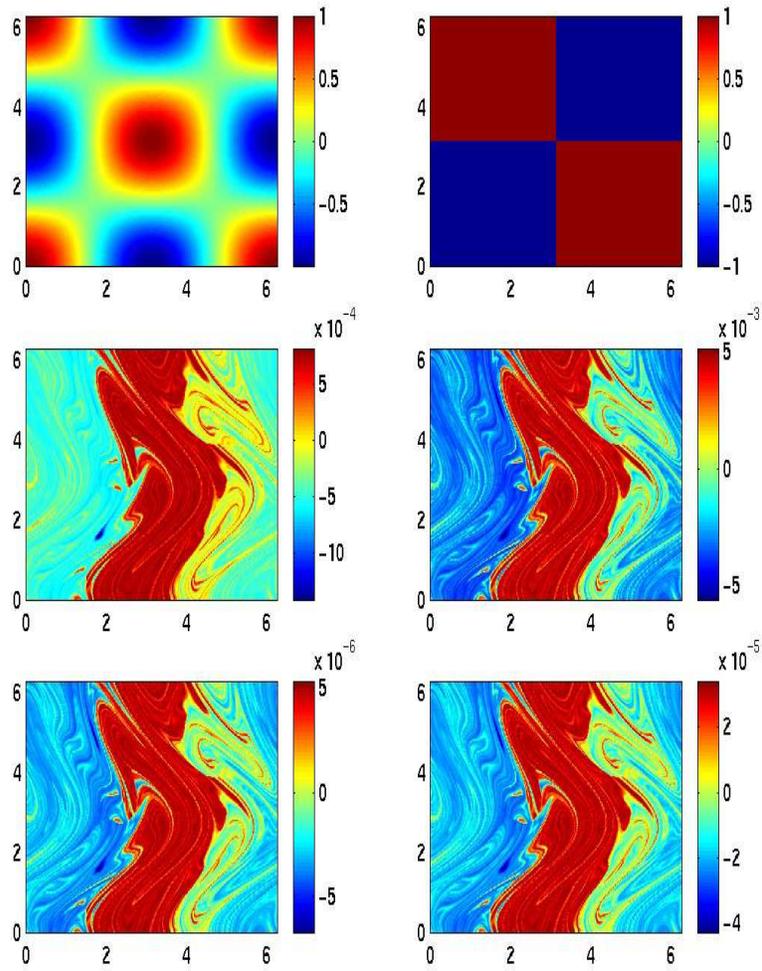}
\end{center}
\caption{The weak barrier case. Upper Panels : The two initial conditions. Middle Panels :
The eigenmodes after 250 iterations. Lower Panels : The eigenmode after
650 iterations.}
\label{fig:fig3}
\end{figure}

\begin{figure}
\begin{center}
\epsfxsize=9.0 cm
\epsfysize=9.0 cm
\leavevmode\epsfbox{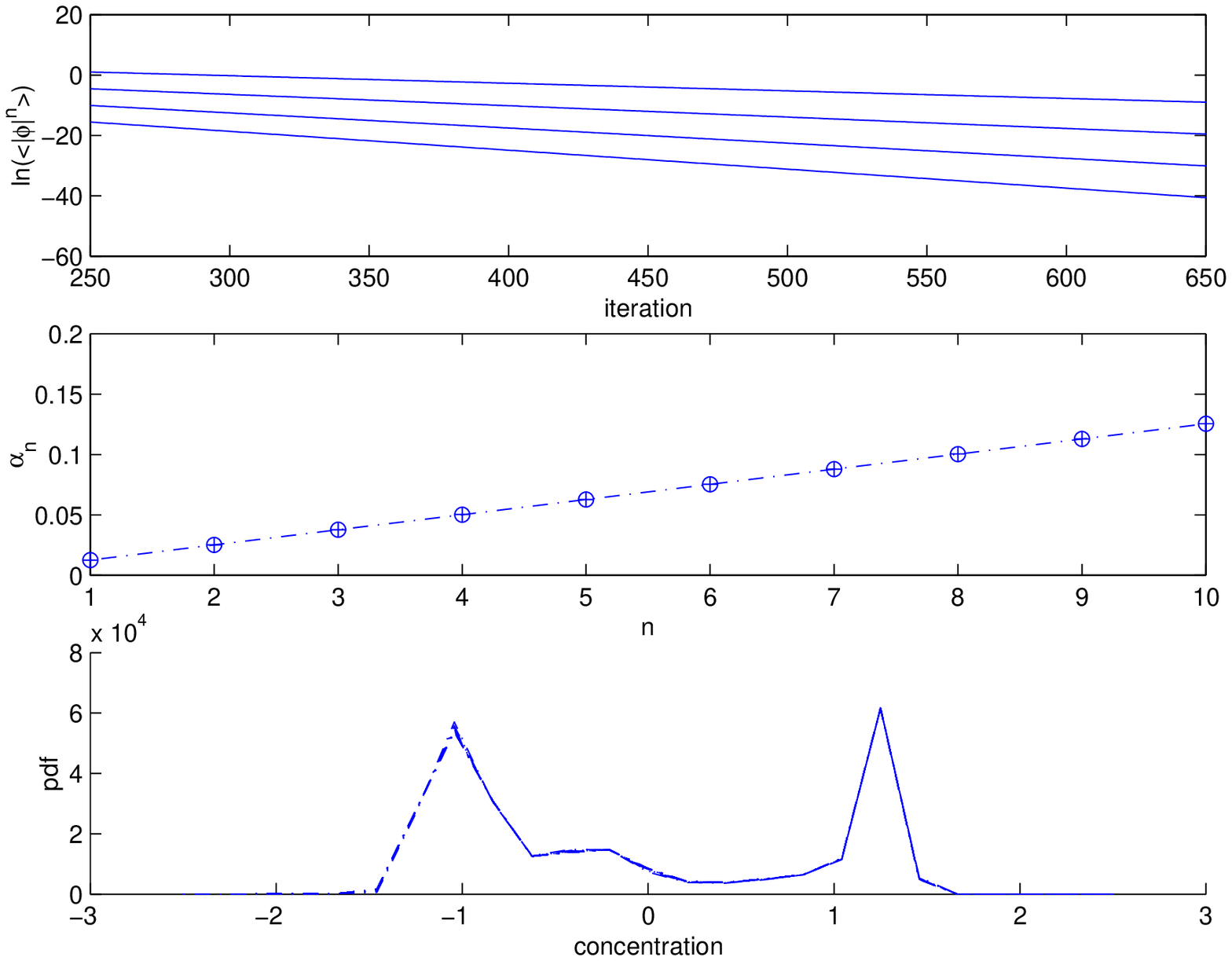}
\end{center}
\caption{The weak barrier case. Upper Panel : The decay of the various moments (n=2,3,4 and 5 with
the higher moments appearing lower on the figure) for the first initial condition.
Middle Panel : The extracted values of $\alpha_n$ Vs. $n$ 
(the dash-dot line marked with circles),
'+' is a plot of $n \alpha_1$. Lower Panel : The pdf's in steps of 25 iterations from iteration 
250 to 650 for IC I. The pdf's lie on top of each other due to their self-similar nature.}
\label{fig:fig2a}
\end{figure}

\begin{figure}
\begin{center}
\epsfxsize=9.0 cm
\epsfysize=9.0 cm
\leavevmode\epsfbox{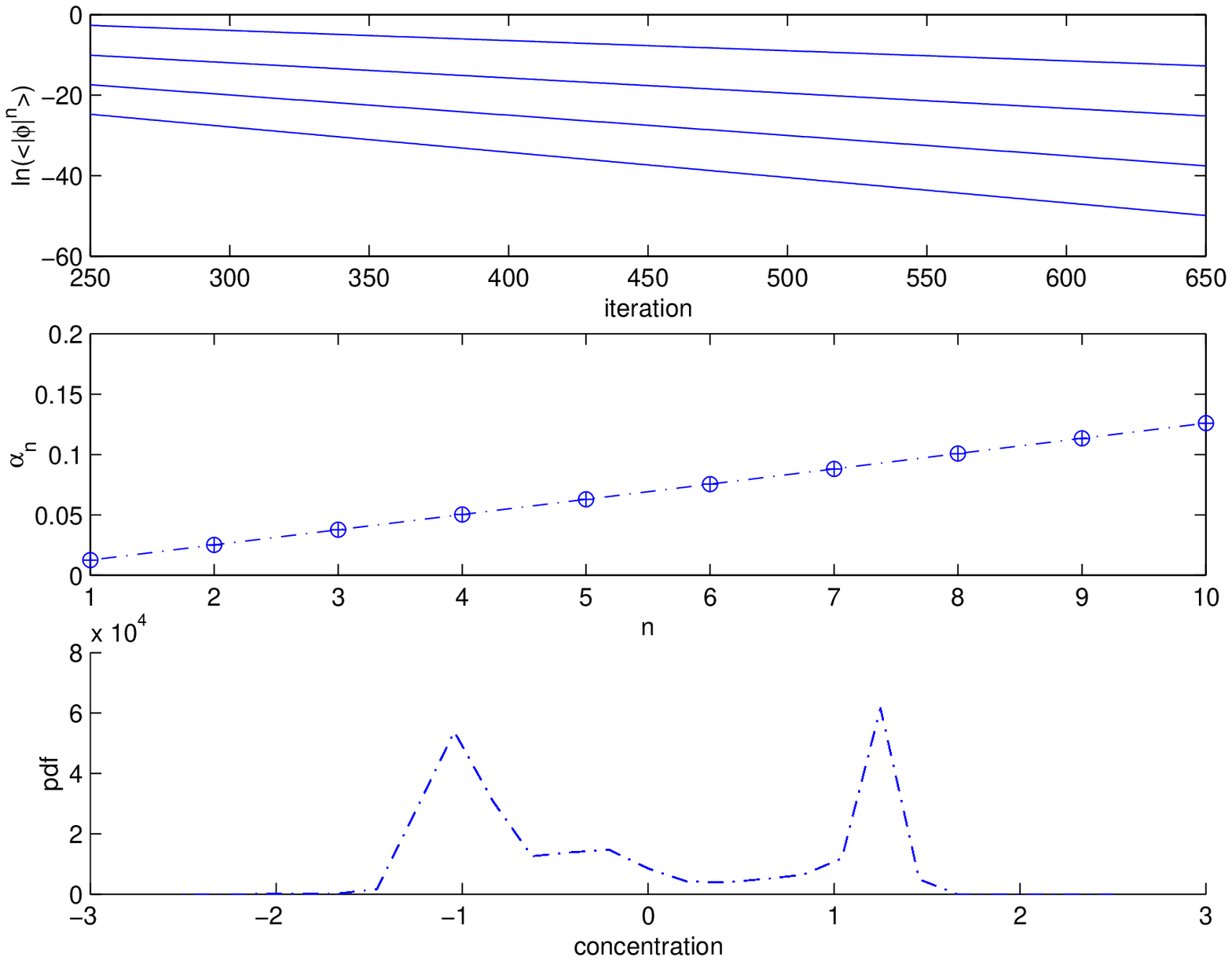}
\end{center}
\caption{The weak barrier case. Upper Panel : The decay of the various moments (n=2,3,4 and 5 with 
the higher moments appearing lower on the figure) for the second initial condition.
Middle Panel : The extracted values of $\alpha_n$ Vs. $n$ 
(the dash-dot line marked with circles),
'+' is a plot of $n \alpha_1$. Lower Panel : The pdf's in steps of 25 iterations from iteration  
250 to 650 for IC II. Once again, the pdf's lie on top of each other due to their self-similarity.}
\label{fig:fig2b}
\end{figure}

On the other hand we get the desired global
decay (down to machine precision) accompanied 
by the emergence of a periodic eigenmode for other pairs of $s_1,s_2$. 
(namely, $s_1 = 5.9698, s_2 = 1.4523$ and $D=0.5$). 
Apart from using $\phi(x,y) = \textrm{cos}(x) \textrm{cos}(y)$ as an initial condition we 
also carried out the simulation using a checkerboard type initial condition where the field
is discontinous but square integrable (the initial conditions can be seen in the upper panel
of Fig. \ref{fig:fig3}).
The spatial structure of the eigenmode at iteration
$250$ and $650$  for the first initial condition can be
seen in the first column of Fig. \ref{fig:fig3}. The second column of Fig. \ref{fig:fig3}
shows the evolution of the scalar field for the second initial condition (the eigenmode is shown
at iteration 250 and iteration 650).
The upper panel Fig. \ref{fig:fig2a} shows a log plot of the moments as a function of iteration.
\footnote{The higher moments involve a fair degree of numerical uncertainly,
even though they do appear to behave as expected,
we feel it is better to rely on the pdf's themselves.\\}
The perfectly normal scaling, i.e.  
$\alpha_n \sim n $, (refer Eq. (\ref{1b})) is shown in the middle panel 
of Fig. \ref{fig:fig2a}. Finally the self similar pdf's in steps of 25 iterations from
iteration 250 to 650 can be seen in the lowermost panel of Fig. \ref{fig:fig2a}.
Fig. \ref{fig:fig2b} shows the same entities for the second initial condition (see figure
caption for details).
The bimodal nature of the pdf indicates that there still exist regions which 
remain relatively isolated from each other. In a sense these regions are "large 
enough" 
to permit chaotic advection of the scalar within themselves and the barriers
are "leaky enough" so 
the exchange of the scalar
across them is strong enough so as {\it not} to be a controlling factor. 
Note that due to the existence of these barriers the value of $D$ has to be
reasonably large (in our experiments the results for $\alpha_n$ were identical for $D>0.2$),
if the system was mixing without any barriers then the results would be valid for 
infinitesimally small diffusivity.
\\

Though the emphasis in the dynamo literature 
has been on the infinite magnetic Reynolds number limit, there have been a few studies 
on the form of the eigenfunctions at finite magnetic Reynolds numbers.
The maps used in these studies 
are chosen for their mathematical properties (such as the Bakers map in 
\cite{Finn-PHF} or 
the Cat map on an unusual Reimannian manifold in \cite{Arn-81}), i.e.,
one has good mixing properties over the whole domain. Also, these maps
allow analytical estimates of the form of the eigenfunctions, 
hence providing some insight into the generically 
singular infinite magnetic Reynolds number limit (in terms of convergence to 
some eigen-distribution \cite{Sow-94}). Unfortunately
such maps are slightly unphysical, their chaotic properties are due to 
boundary conditions or due to the curvature of the underlying manifold 
rather than any inherent nonlinearity in the maps themselves. 
Even though we choose not to use such maps for advective purposes, we would like to mention that
their results (especially \cite{Finn-PHF}) are consistent with our work. 
\\

\subsection{Random maps (Aperiodic Velocity Fields)}
\begin{figure}
\begin{center}
\epsfxsize=9.0 cm
\epsfysize=9.0 cm
\leavevmode\epsfbox{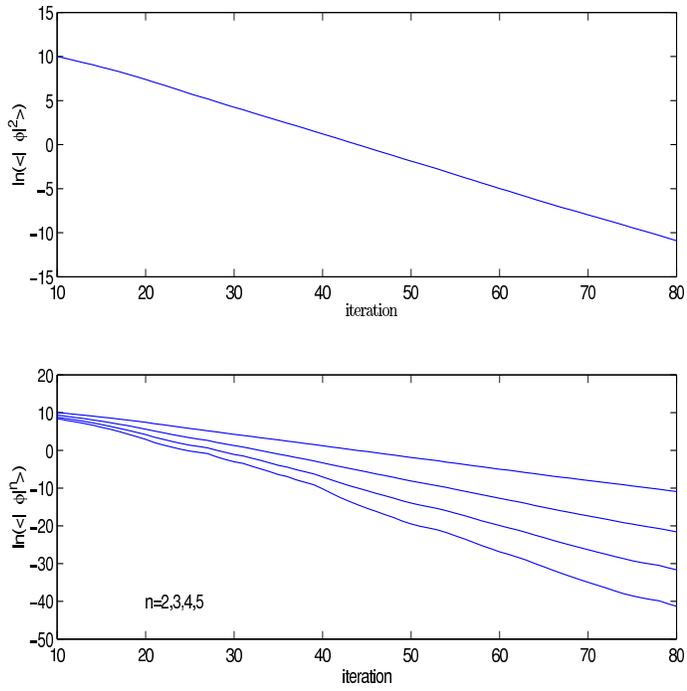}
\end{center}
\caption{The aperiodic (no barrier) case. Upper Panel : The decay of the variance.
Lower Panel : The higher order moments (progressively higher moments appear lower on the figure).}
\label{fig:fig5}
\end{figure}

\begin{figure}
\begin{center}
\epsfxsize=9.0 cm
\epsfysize=9.0 cm
\leavevmode\epsfbox{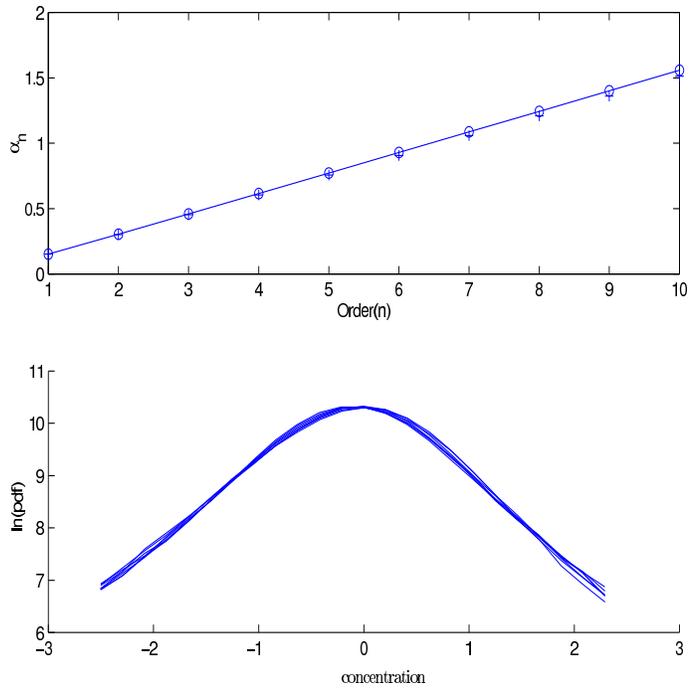}
\end{center}
\caption{The aperiodic (i.e., no barrier) case. Upper Panel : $\alpha_n$ Vs. $n$ (o), the dash dot
line is $n\alpha_1$. Lower Panel : The self similar pdf's. }
\label{fig:fig6}
\end{figure}

In the aperiodic case there is no clean way to separate time from space 
as was done in the earlier situations. Due to this we weaken the notion of an 
eigenfunction by invoking statistical stationarity, i.e., the scalar field is
said to enter an eigenmode when the shape of its pdf remains invariant with time
(upon renormalization by its variance as the overall field
strength is decaying). Note that the earlier classical 
eigenmode also has self-similar pdf's. Of course it has a stronger form of convergence
where the scalar field itself approaches a stationary (or periodic) spatial structure. 
The random map used to represent the aperiodic flow 
is again a nonlinear shear flow,

\begin{eqnarray}
x_{n+1} = x_{n} + 4 ~ \textrm{sin}(y_{n} + p_n) \quad  \nonumber \\
y_{n+1} = y_{n} + \textrm{sin}(x_{n+1} + q_n) \quad 
\label{1j}
\end{eqnarray}
Now $p_n , q_n$ ($\in [0,2\pi]$) are random numbers which are selected 
at the beginning of each iteration. 
$x_{n},y_{n}$ are $\textrm{mod}(0,2\pi)$, $D=0.25$ and the initial scalar field is 
$\phi(x,y) = \textrm{cos}(0.5x) \textrm{cos}(0.5y)$. The map is iterated till the variance goes down
to machine precision. The variance as a function of time can be seen in the upper panel of
Fig. \ref{fig:fig5}. After an initial transient (which lasts for about $10$ iterations,
see \cite{Ant})
the variance decays 
exponentially. The higher moments can be seen in the lower panel of
Fig. \ref{fig:fig5}. The non-anomalous nature of the exponents can be seen in the 
upper panel of Fig. \ref{fig:fig6}, the implied self-similar behaviour of the pdf of the 
scalar field is demonstrated in the lower panel of Fig. \ref{fig:fig6}. The pdf is 
unimodal as there are no isolated regions in the domain (the randomness at each iteration is 
responsible for the destruction of any barriers that might exist in the steady map). 
Moreover, the shape of the
pdf is characterized by a Gaussian core and stretched exponential tails, as per the 
higher resolution studies in Pierrehumbert \cite{Ray-Chaos}.
\\

\subsection{Physical Interpretation}
The physical picture that goes along with the eigenmode is as follows : 
Consider a filament of scalar whose length is comparable to the scale of 
variation of the velocity field (or the typical size of an eddy in the velocity
field). Due to the assumed chaotic properties of a generic time dependent 2D 
flow, the filament will tend to be stretched out. The point is that,
as the scale of the filament is already the same as that of the flow, 
instead of being merely stretched
the filament will fold and start to fill the eddy. This process 
continues till the filament 
has been "packed" as tightly possible. Note that it is the
diffusion that is responsible for the "packing", i.e., for $\kappa >0$ there is
a limit on how thin a filament can get.
Once this situation is reached, as
the problem is unforced, the whole structure remains stationary (or periodic or 
statistically stationary depending on the flow) till the diffusion destroys all
the variance in the scalar field. We argue that the eigenmode is a representation
of this "packed" structure.
\\

\section{The Different Regimes}

Having gotten a feel for the situation when the initial scale of variation 
of the scalar field and that of the velocity field are comparable, we proceed to 
look into the evolution of an initially small scale scalar field, i.e., 
$l_{\kappa} << l_{s} << l_{v} \sim L$.
\\

As $\phi \in {\it B}(D)$, via Fourier's theorem, we can represent $\phi$ as,

\begin{equation}
\phi(\vec{r},t) = \int_{\vec{k_0}} \hat{\phi}(\vec{k_0},t) ~ e^{i\vec{k}(t)\cdot\vec{r}} d\vec{k_0}
\label{2a}
\end{equation}
Substituting a plane wave solution in Eq. (\ref{1a}) and equating the real and imaginary parts, we
get, 

\begin{equation}
\frac{\partial \hat{\phi}(\vec{k_0},t)}{\partial t} =  -\kappa ~ {|\vec{k}(t)|}^2 ~ \hat{\phi}(\vec{k_0},t)
\label{2b}
\end{equation}
where $\vec{k}(t)$ is given by,

\begin{equation}
\vec{r} \cdot \frac{\partial \vec{k}(t)}{\partial t} = - \vec{u} \cdot ~ \vec{k}(t) ; \quad \vec{k}(0)=\vec{k_0}
\label{2c}
\end{equation}
\\

\subsection{Small time scales}

In the eigenvalue formulation, when $l_{s} << L \sim l_{v}$ the scalar field is 
essentially in an infinite domain, the implication is that the spectrum of ${\mathcal L}_{\kappa}$
now typically consists of a continuous part and hence possesses eigenvalues which lie 
arbitrarily close to each other. As there is no dominant eigenvalue we will not see the
emergence (at short times) of the corresponding eigenmode. 
\\

Taking advantage of $l_s << l_v$ we linearize Eq. (\ref{2c}) to obtain, 

\begin{equation}
\frac{\partial \vec{k}(t)}{\partial t} = - {\nabla \vec{u}}(\vec{k_0},t) ~ \vec{k}(t)
\label{2d}
\end{equation}
From Eq. (\ref{2b}) we have,

\begin{equation}
\hat{\phi}(\vec{k_0},t) = \hat{\phi}(\vec{k_0},0) ~ e^{ -\kappa {\int_0}^{t} {|\vec{k}(s)|}^2 ~ ds}
\label{2f}
\end{equation}
Therefore, in order to see how the moments behave at small times we need an estimate of how
$|\vec{k}(t)|$ evolves. Generally this is a fairly intricate question as the
solution to Eq. (\ref{2d}) involves a time ordered exponentiation which 
reduces to 
a product of matrices \cite{Cri-book}. 
\\

\subsubsection{Useful properties of the FTLE's}

We mention some of the properties of the FTLE's that will be needed for calculations
in the forthcoming sections. In $2D$ we have the possibility of $2$ (asymptotic) Lyapunov 
exponents. Let us denote the larger of these by $\Lambda_0$. Furthermore let us
denote the 
FTLE along a particular trajectory, after $n$ iterations, by $\Lambda(n)$.
Note that physically the FTLE's are defined with respect to the change in volume of
a given set of initial conditions. In $2D$ incompressible systems this only depends
on the larger eigenvalue in the above mentioned matrix product. In higher
dimensions the situation is more complicated as the FTLE depends on the sum of the positive eigenvalues
(for continuous flows) of the matrix product, i.e, it is closely related to the topological entropy
of the dynamical system.
\\

In $D$ dimensions one has the possibility of $D$ (asymptotic) Lyapunov exponents. Let us
denote them by $\Lambda_i ~(i=1:D)$.
It is generally assumed that, along a particular trajectory, the probability of $\Lambda_i(n)$
(i.e., the spectrum of Lyapunov exponents calculated after $n$ iterations) 
deviating from $\Lambda_i$ decays 
exponentially with $n$ \cite{Eck-86}. In fact it is
generally taken for granted that each $\Lambda_i(n)$ is governed by the central limit
theorem (CLT) and is distributed around $\Lambda_i$ \cite{Grass-88}
(see also the discussion in Section $9.4$ of \cite{Ott}). 
We would like to 
mention that we are not aware of a general proof of this statement. Having said this,
we mention that the results of Balkovsky and Fouxon \cite{BF-99} appear to 
lead in this direction though the status of the proof of the general statement is not very clear. 
The 
specific results we are aware of start with the work of Furstenberg (see \cite{Cri-book}),
who showed that if the matrices 
are i.i.d (i.e. delta correlated in time) random matrices
and if the system is incompressible, then $\Lambda_0 (=\textrm{max}(\Lambda_i) > 0$. 
A stronger result, which is
valid only in $2D$, has recently been shown by
Chertkov et. al. \cite{Chert-96}. It states that if the tangent maps are random matrices
with arbitrary (but finite) correlation time then $\Lambda_0 > 0$, moreover $\Lambda(n)$
obeys the CLT and is distributed around $\Lambda_0$. 
\\

Therefore, using the CLT for $\Lambda(n)$ in $2D$, the distribution of the FTLE's can be
expressed as (see Section $8.6.4$ of \cite{Frisch} for the large deviation result),

\begin{equation}
Q(\Lambda,n) \sim e^{-n G(\Lambda - \Lambda_0)}
\label{2f4}
\end{equation}
where $G(\Lambda - \Lambda_0)$ is the Cramer function and $\Lambda_0$ is the asymptotic
Lyapunov exponent. Due to the mixing nature of the system, as $n \rightarrow \infty$ 
the FTLE's along (almost) every trajectory will tend 
to $\Lambda_0$, 
moreover, the
average of many realizations along a particular trajectory can be interpreted as a spatial
average \cite{Aref-91}. 
\\

\subsubsection{The scalar pdf}

At this initial stage the problem is completely reversible in the sense that the area of a given
blob of scalar is invariant. 
Physically when we release a small blob of scalar in a chaotic flow the tendency is for the
blob to form a filament while respecting the conservation of area. 
Mathematically this implies that on average $\vec{k}(t)$ grows with
time, i.e., one of the components of $\vec{k}(t)$ increases while the other decreases. 
The inference from Eq. (\ref{2f}) is that the moments of the scalar field
decay in a faster than exponential fashion (see for eg. \cite{Zel-84}, \cite{Elp}).
\\

Keeping in mind
that the backwards in time problem has the same FTLE's as the forward problem, we can
envision a blob of scalar at a time $t$ to have resulted from the diffusive homogenization
of a filament (at $t=0$) which has undergone advective collapse \cite{Ray-Chaos}. 
As the initial scale of variation of the scalar field is very small, in effect the blob
is a result of the diffusive homogenization of a large number of independent random 
concentrations. Denoting the scale of the blob by $l$, for one realization, the probability
of the blob's concentration being $\phi = \phi_1$ at time $t$ is \cite{Ray-Chaos}, 

\begin{equation}
P(\phi_1)_t \sim e^{-\frac{l e^{\Lambda t}}{l_s} S(\phi_1)}
\label{2h}
\end{equation}
$S(\phi)$ is the Cramer function. For notational simplicity we have taken the mean of 
the random concentrations making up the blob to be $0$. Note that the chaotic nature of the
flow is embedded in the exponential prefactor to $S(\phi)$. For many realizations the 
average probability of the above event is,

\begin{equation}
<P(\phi_1)_t> \sim \int e^{-\frac{l e^{\Lambda t}}{l_s} S(\phi_1) - tG(\Lambda - \Lambda_0)} ~ d\Lambda
\label{2i}
\end{equation}
Here $<\cdot>$ represents an average of many realizations along a particular trajectory. 
As mentioned, this can be interpreted as a spatial average.
By a steepest descent argument the value of $\Lambda$ that dominates the above integral
satisfies (denoting it by $\Lambda_1$),

\begin{equation}
- \frac{l}{l_s} S(\phi_1) e^{\Lambda_1 t} = G'(\Lambda_1 - \Lambda_0)
\label{2j}
\end{equation}
As $S(\phi) \ge 0$ Eq. (\ref{2j}) implies $\Lambda_1 \le \Lambda_0$. 
Substituting in Eq. (\ref{2i}) we have,

\begin{equation}
<P(\phi_1)_t> \sim  e^{G'(\Lambda_1 - \Lambda_0)} e^{-\beta t} ~;~ \beta = G(\Lambda_1 - \Lambda_0)
\label{2j1}
\end{equation}
Therefore the probability of a particular deviation 
decays exponentially with a rate $\beta$.
The important point is that $\beta$ is a function of the deviation through $\Lambda_1$. 
As $S(\phi)$ is convex \cite{Frisch}, for
$\phi = \phi_2$ (where $|\phi_2| > |\phi_1|$), $S(\phi_2) > S(\phi_1)$. From Eq. (\ref{2j}) the 
dominant FTLE for $\phi_2$ (denoting it by $\Lambda_2$) satisfies, $\Lambda_2 < \Lambda_1$.
Therefore the probability of a large deviation is governed by the smaller FTLE's, i.e., the tails of 
the pdf of $\phi$ are sensitive to the tails of the FTLE distribution. At the other extreme,
as $\phi \rightarrow 0$ we have $\Lambda_1 \rightarrow \Lambda_0$ (because $S(\phi) \rightarrow 0$ as
$\phi \rightarrow 0$ and $G'(\Lambda - \Lambda_0) \rightarrow 0$ as $\Lambda \rightarrow \Lambda_0$). 
Also, $S(\phi) \sim \phi^2$ for small deviations. From Eq. (\ref{2h}) the implication is
that, in this regime, the scalar pdf has a Gaussian core with a width that decreases in time.
\\

\subsubsection{Numerical results}
\begin{figure}
\begin{center}
\epsfxsize=9.0 cm
\epsfysize=9.0 cm
\leavevmode\epsfbox{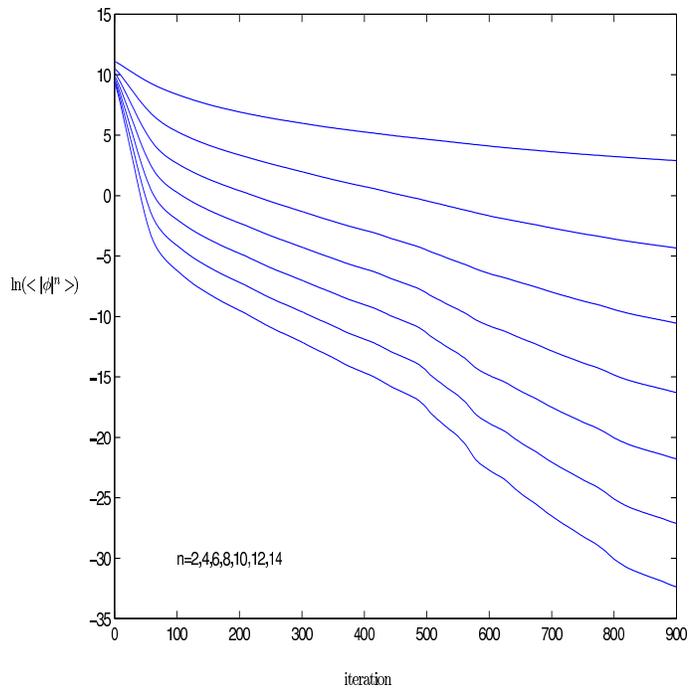}
\end{center}
\caption{The decay of the moments for a scalar field whose initial scale of variation
is small as compared to that of the advecting flow (the higher moments are lower on the
figure).}
\label{fig:fig7}
\end{figure}

\begin{figure}
\begin{center}
\epsfxsize=9.0 cm
\epsfysize=9.0 cm
\leavevmode\epsfbox{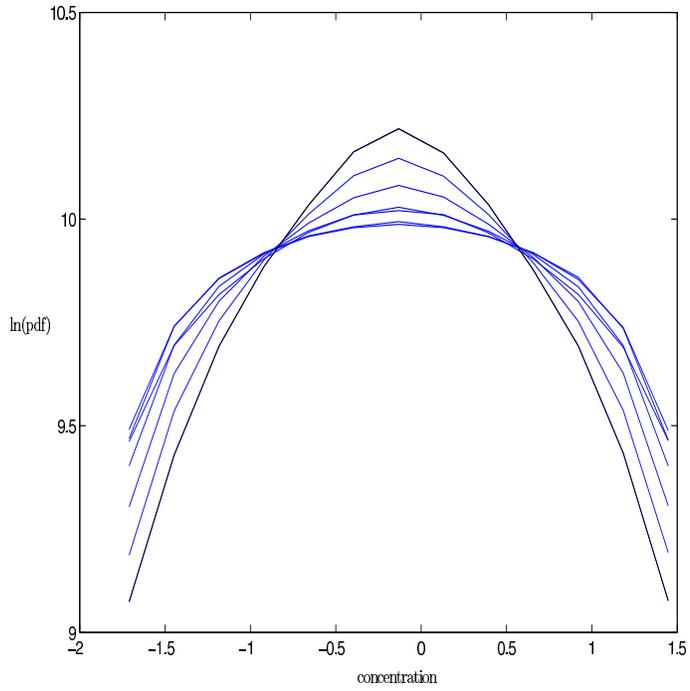}
\end{center}
\caption{The pdf's from iteration $40$ to $100$ (bold line at iteration $100$) for the initially
small scale scalar field.}
\label{fig:fig8}
\end{figure}

To numerically verify these predictions we use the map,

\begin{eqnarray}
x_{n+1} = x_{n} + A ~ \textrm{sin}(B y_n + {\psi}_n) \quad \textrm{mod}(0,12\pi) \nonumber \\
y_{n+1} = y_{n} + C ~ \textrm{sin}(D_1 x_{n+1} + {\sigma}_n) \quad \textrm{mod}(0,12\pi)
\label{2g}
\end{eqnarray}
${\psi}_n, {\sigma}_n$ ($\in [0,2\pi]$) are random numbers chosen at the beginning
of each iteration, $B=D_1=\frac{1}{6}$ so as to make the scale of the flow comparable to the
scale of the domain and $A=1,C=1$. Diffusion is represented as in Eq. (\ref{1h}) with $D=0.25$.
To satisfy $l_s << l_v$ we take the initial scalar field to be
$\phi(x,y) = \textrm{cos}(4x) \textrm{cos}(4y)$. Fig. \ref{fig:fig7} shows the evolution of
the moments for an ensemble average over $25$ realizations of the map. The faster than
exponential decay of the moments is visible for the first 90 to 100 iterations.
Moreover, the pdf evolution (in steps of 10 iterations) seen in Fig. \ref{fig:fig8} 
goes along with theoretical expectations.
\\

\subsection{Intermediate time scales}

In Fig. \ref{fig:fig7} there is a transition from faster than exponential
to purely exponential
decay of the moments at around iteration $100$.
This transition is a diffusive effect (the diffusive time $T$ in the introduction refers to
the time at which this transition occurs). As mentioned,
from Eq. (\ref{2f}) it is evident that the growing component of $\vec{k}(t)$ controls
the decay of $\phi(\vec{k}_0,t)$.
In physical space the growing component of $\vec{k}(t)$ represents a shrinking physical
scale. As there is a lower limit to this shrinking scale (i.e. $l_{\kappa}$), the
magnitude of $\vec{k}(t)$ saturates when its growing component becomes comparable
to ${l_{\kappa}}^{-1}$. From this time onwards ${|\vec{k}(t)|}^2$
fluctuates around ${l_{\kappa}}^{-2}$ and hence, $\phi(\vec{k}_0,t)$ decays in a
purely exponential fashion. Physically, even though the flow is
incompressible, due to the nonzero diffusivity
the area of a blob of scalar is no longer invariant. In fact, the area increases exponentially, 
implying
an exponential decay of the scalar concentration within the blob itself. 
\\

\subsubsection{Existing results on $\alpha_n$}
In this regime, one where both $l_{s} << L \sim l_{v}$ and
the area of a given blob of scalar evolves, 
the moments of the scalar field have been dealt with in detail by Balkovsky and
Fouxon \cite{BF-99}. Their approach involves shifting to a comoving
reference frame and using the effective equation \cite{BF-99},\cite{Chert-96},

\begin{equation}
\frac{\partial \phi}{\partial t} + \sigma_{\alpha \beta} ~ r_\beta \nabla_\alpha \phi = \kappa \nabla^2 \phi
\label{2k}
\end{equation}
The essential point being that now the trajectory equation takes the form,

\begin{equation}
\frac{\partial r_\beta}{\partial t} = \sigma_{\alpha \beta}(t) ~ r_\beta
\label{2l}
\end{equation}
For this problem the Lyapunov exponents are defined as (see for eg. \cite{Chert-96}),

\begin{equation}
\lambda(t) = \frac{1}{t} \textrm{log} \frac{|\vec{r}(t)|}{|\vec{r}(0)|}
\label{2m}
\end{equation}
Due to the effective space time separation of the velocity
field in the comoving reference frame, $\lambda(t)$ is only a function of time. In effect the
behaviour of $\Lambda(n)$ for a given trajectory in the FTLE formalism is now valid for
the entire domain.
In fact, the aforementioned results of Chertkov et. al. \cite{Chert-96} are actually shown in the 
comoving framework for $\lambda(t)$.
Building on this framework,
Balkovsky and Fouxon \cite{BF-99} derive expressions for the
moments of the scalar field by considering the evolution of an inertia tensor
like quantity for a blob of scalar.
The advantage of the inertia tensor like quantity is that its evolution
explicitly captures the exponential growth of a blob's area which characterizes this 
regime.
The chief assumption is that the eigenvalues of the inertia tensor, at any finite time $t$,
are governed by the
CLT and are distributed about
the asymptotic Lyapunov exponents $\lambda_i (i=1,2 ~ \textrm{in}~ 2D)$. 
With this they are able to show that $<|\phi(t)|^n> \sim e^{-\alpha_n t}$ and that $\alpha_n$
is a nonlinear function of $n$, implying a {\it non} self similar pdf for the scalar field. 
\\

\subsubsection{A simple example of a single non-self overlapping blob.}
We present a simple example to put some of the notions regarding the behaviour of $\alpha_n$ in
perspective. The objective is to analyze the evolution in concentration of a single non-self 
overlapping \footnote{ What we mean by "non-self overlapping" is a single blob that stretches
and folds in this large scale velocity field but the filaments of the blob do not overlap
with each other in these intermediate time scales.}
blob. 
Let the
blob's concentration be $C_{T}$ (at time $t=T$). Now, the 
increase in area of the blob is controlled by the FTLE at the position of the blob. 
Therefore we have,

\begin{equation}
C_t = C_{T} ~ e^{- \Lambda (t-T)} ~;~ t \ge T
\label{2n}
\end{equation}
Hence,

\begin{equation}
<|C_t|^n> = |C_{T}|^n ~ \int e^{ - n\Lambda (t-T)}  Q(\Lambda,t) ~ d\Lambda
\label{2o}
\end{equation}
where $<\cdot>$, as before, indicates an average of many realizations over one trajectory.
Therefore we have,

\begin{equation}
<|C_t|^n> = |C_{T}|^n ~ \int e^{ - n\Lambda (t-T) - tG(\Lambda - \Lambda_0)}  ~ d\Lambda
\label{2p}
\end{equation}
By a steepest descent argument, for $t>>T$, the above integral is dominated for $\Lambda =
\Lambda^*$ (implicitly) given by,

\begin{equation}
\frac{dG}{d\Lambda}|_{\Lambda=\Lambda^*} = -n
\label{2p1}
\end{equation}
This implies $\Lambda^* < \Lambda_0$, moreover for small $n$ Eq. (\ref{2p1}) implies 
$\Lambda^* \rightarrow \Lambda_0$. In this region the Cramer function is parabolic,
i.e., taking $G(\Lambda - \Lambda_0) = a(\Lambda - \Lambda_0)^2$ we have,
 
\begin{equation}
\Lambda^* = \Lambda_0 - \frac{n}{2a}
\label{2q}
\end{equation}
Substituting back we obtain (for small $n$),

\begin{equation}
<|C_t|^n> \sim e^{-\alpha_n t} ~;~ \alpha_n = n(\Lambda_0 - \frac{n}{4a})
\label{2r}
\end{equation}
Immediately it is clear that the pdf
of the scalar field will be {\it non} self similar as $\alpha_n$ is a nonlinear function
of $n$. Furthermore, the nonlinearity is due to fact that there is a distribution of
FTLE's. If we had a single FTLE or if the FTLE distribution collapsed to a delta function 
in a short time then the scaling would be non anomalous with $\alpha_n = n \Lambda_0$.
Note that the explicit expression for $\alpha_n$ given in Eq. (\ref{2r}) is only 
valid for small $n$. 
From Eq. (\ref{2p1}) the higher moments (as in the previous regime) are sensitive to the smaller FTLE's.
As our simple example follows Balkovsky and Fouxon's work, their form for 
$\alpha_n$ (through more detailed considerations which take into account the 
diffusive overlap of many blobs) is similar to Eq. (\ref{2r}) (see Eq. 3.9 in 
\cite{BF-99}). 
We reiterate that even though the FTLE distribution is responsible for the pdf
evolution in this regime, fundamentally this is still a diffusive effect (in the sense that
it is the nonzero diffusivity that leads to the area evolution which characterizes this regime).
\\

\subsubsection{Numerical results}

\begin{figure}
\begin{center}
\epsfxsize=9.0 cm
\epsfysize=9.0 cm
\leavevmode\epsfbox{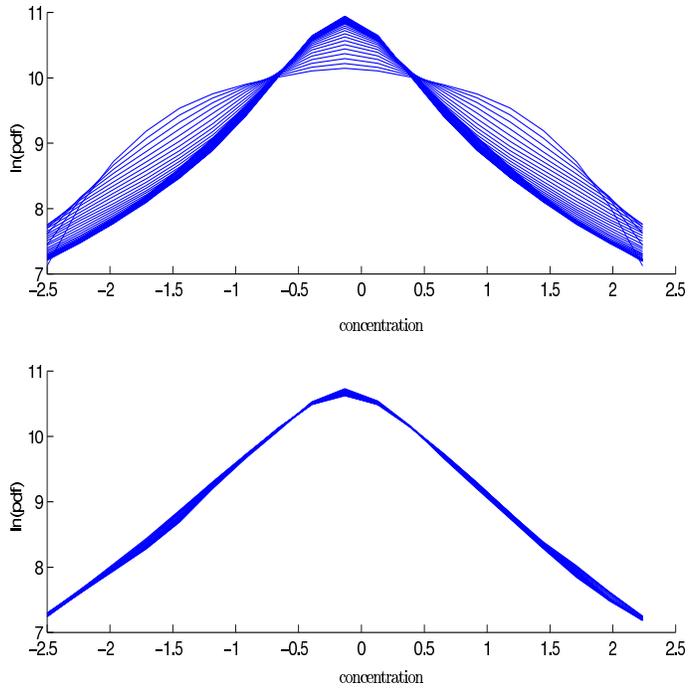}
\end{center}
\caption{Initailly small scale scalar field. Upper panel : pdf's from $90 - 490$ (non
self similar). Lower panel : pdf's from $600 - 800$ (self similar).}
\label{fig:fig9}
\end{figure}

\begin{figure}
\begin{center}
\epsfxsize=9.0 cm
\epsfysize=9.0 cm
\leavevmode\epsfbox{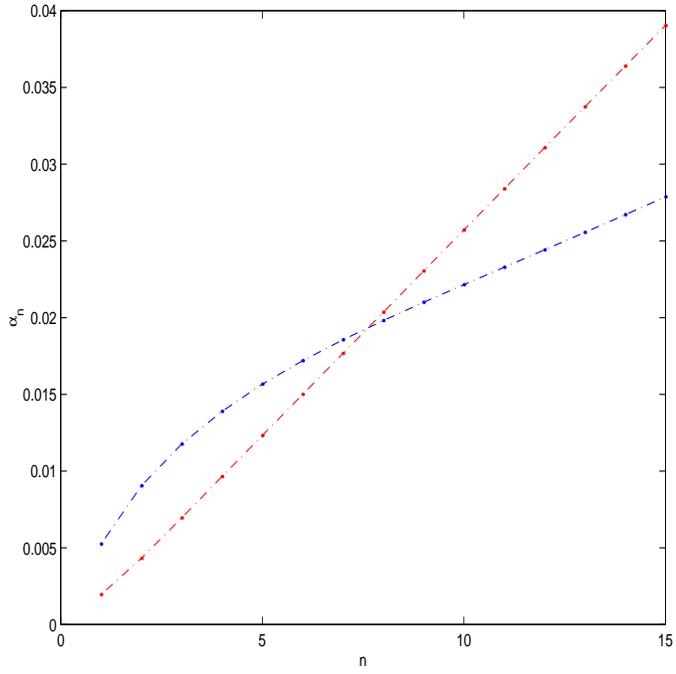}
\end{center}
\caption{Initially small scale scalar field : $\alpha_n$ Vs. $n$
(blue (nonlinear) - iteration $90 - 490$, red (linear) - iteration $600 - 800$).}
\label{fig:fig10}
\end{figure}

The upshot of these considerations is that at these intermediate time scales we should observe
an evolution in the pdf of the scalar field.
To verify this, 
we plot the pdf's (in steps of $10$ iterations from iteration $90$ to iteration $490$) 
in the upper panel of 
Fig. \ref{fig:fig9}. The {\it non} self similarity is clearly evident.
The anomalous nature of $\alpha_n$ can be seen in Fig. \ref{fig:fig10}. 
We observe that the pdf's are 
characterized by a (progressively smaller) Gaussian core and (progressively
fatter) stretched exponential tails. Interestingly after about iteration $400$
the tails of the pdf tend to relax back to a purely exponential form (which is the 
theoretically expected shape, see section IIIB in Falkovich et. al. \cite{Falk}). The only 
discrepancy
with theoretical predictions \cite{BF-99} is that $\alpha_n$ does not saturate at high $n$, this
might be due to the fact that we are numerically unable to get to high enough $n$ to
actually observe the saturation ({\it note} : see the example in the Discussion).
\\

\subsection{Large time scales}
From the preceeding arguments it would appear that the scalar pdf
is an ever evolving entity (as is implied in \cite{BF-99}), but
once again from Fig. \ref{fig:fig7} we see that the behaviour of the moments undergoes
another change between iterations $500-600$. We claim that it is at this stage that the scalar
field actually enters the eigenmode, i.e, beyond iteration $600$ the 
pdf should become self similar. The lower panel of Fig. \ref{fig:fig9} shows the pdf's
(again in steps of $10$ iterations from iteration $600$ to iteration $800$). The self
similar nature of the pdf's is clearly evident, also $\alpha_n$ (shown in 
Fig. \ref{fig:fig10} along with the $\alpha_n$ from the previous regime) follows
$\alpha_n \sim n$.
Physically the scale of variation of the scalar field 
is now comparable to that of the velocity field, hence the velocity field separation
used to derive Eq. (\ref{2k}) is invalidated. 
Correspondingly in the FTLE formulation the
linearization breaks down.
In fact, the problem is now in a situation 
where the mathematical considerations leading to the eigenmode (Section II) become applicable.
\\

\section{Discussion}
In the first part of the paper we described the behaviour of a scalar field whose 
initial scale of variation was comparable to that of the advecting flow (both of which were
similar to the size of the domain).
In both, periodic and
aperiodic situations we demonstrated the emergence of scalar eigenmodes. In spite of the weak 
barriers that exist in the deterministic map used for periodic flows, the emergence of a 
scalar eigenmode is robust (as is seen from the two different initial conditions). 
In the case of random maps, representing aperiodic flows, 
the mixing is global (as the barriers are destroyed) 
and the eigenmode is statistical in nature. 
Along with the recent work of Fereday et. al. \cite{Fer} this indicates that 
passive scalar advection-diffusion in chaotic flows can be insightfully treated as 
eigenvalue problem as is commonly done for steady flows (see for eg. Young et. al. \cite{Young}).
\\

We then looked at the evolution of a passive
scalar whose initial scale of variation was small as compared to that of the advecting flow
(again, the scale of the flow was comparable to the size of the domain).
Initially, as has been noted in previous studies \cite{Zel-84}, \cite{Elp}, 
the moments of the scalar field decay in a faster than exponential
fashion. The pdf of the scalar field in this
initial regime is an evolving entity, it is shown to be characterized by a Gaussian core which 
shrinks with time. When the scalar filaments 
reach the diffusive scale the behaviour of the moments 
experiences a transition to a purely exponential decay. We emphasize that this
transition is a diffusive effect. In this intermediate regime the pdf of the scalar field 
is still an evolving entity (as is shown by Balkovsky and Fouxon \cite{BF-99}, via the 
nonlinear dependence of $\alpha_n$ on $n$). By means of a simple example, the nonlinear nature 
of $\alpha_n$ is shown
to be dependent on the distribution of FTLE's, moreover, the higher moments are 
seen to be sensitive to the tails 
of the FTLE distribution. Finally, when the scalar filamants have stretched and 
folded to fill the domain (or have been "packed" as per our previous nomenclature) 
the field enters an eigenmode. The eigenmode, as before, is characterized by 
self similar pdf's. Numerical results of advection on a lattice
followed by diffusion appear to confirm these stages of evolution.
\\

\begin{figure}
\begin{center}
\epsfxsize=9.0 cm
\epsfysize=12.0 cm
\leavevmode\epsfbox{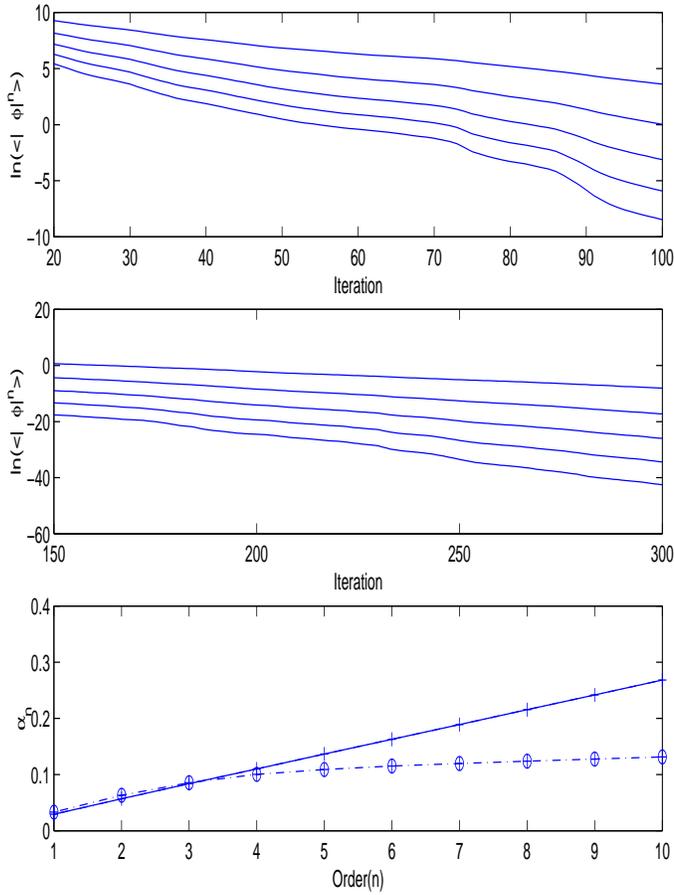}
\end{center}
\caption{Case with A=C=5 in Eq. (\ref{2g}). Upper two panels show the moments (n=2,4,6,8 and 10 with 
higher moments appearing lower on the figure) and the lower
panel shows $\alpha_n$ Vs. $n$ extracted from iterations
20-70 (dashed line) and iterations 150-300 (solid line) respectively. Note the saturation of 
$\alpha_n$ in the intermediate regime.}
\label{fig:fig11}
\end{figure}

The above mentioned stages that 
an initially small scale scalar field encounters are robust, i.e.,
once one enters a particular stage, the statistical 
properties of the scalar field are fixed. A caveat is that the duration of the stages,
strongly depends on the strength of the advecting flow.  As an illustration,
consider the same scalar field and map as
in Eq. (\ref{2g}), but set $A=C=5$ (i.e., a flow with stronger stretching properties). 
The behaviour of the moments and $\alpha_n$ is shown in 
Fig. \ref{fig:fig11} (note that in this case, for the intermediate regime, $\alpha_n$ does
appear to saturate for large $n$). The stronger nature of the flow causes the first stage to 
be very short, the intermediate stage is also relatively shorter as the filaments fill the
domain quickly. Finally, the eigenmode is realized and persists till all the variance is 
destroyed.
\\

In essence the picture that emerges is fairly straightforward, as long as there is a
valid scale separation (i.e., till $l_s << l_v \sim L$) the pdf's of the scalar field are 
evolving entities, albeit involving stages characterized by distinct decay of moments. 
Whereas, as soon as $l_s \sim l_v \sim L$, the scalar field enters
an eigenmode with stationary or self similar pdf's. 
Interestingly, some of our preliminary
numerical work indicates that when we consider $l_v << l_s \sim L$, the self similarity
is maintained whereas the shape of the pdf is altered. In fact, smaller the scale of the
velocity field the more Gaussian is the self similar scalar pdf.
As future work we plan to look deeper into these cases, especially in light of the 
recent results on non diffusive asymptotic self similarity of turbulent decay (i.e. where the 
advecting flows themselves are self similar and nonsmooth) as put forth
by Chaves et. al. \cite{Chaves}.  
\\

\acknowledgments

We are indebted to Prof. Peter Haynes for many valuable discussions and for providing 
access to an early version of the work by Feredey et. al. \cite{Fer}, which led us
to think of the line of research reported above. This work was supported by 
the National Science Foundation, under grants ATM-9505190 and ATM-0123999.

\clearpage

\clearpage

\clearpage

\clearpage

\clearpage

\clearpage

\clearpage

\clearpage

\clearpage

\clearpage

\clearpage

\end{document}